\documentclass[aps]{revtex4}
\usepackage{graphicx}
\usepackage[all]{xy}
\usepackage{amsmath}
\usepackage{amssymb}
\usepackage{color}

\newcommand{\bb}{\bibitem}
\newcommand{\bes}{\begin{subequations}}
\newcommand{\ees}{\end{subequations}}
\newcommand{\benn}{\begin{eqnarray*}}
\newcommand{\eenn}{\end{eqnarray*}}
\newcommand{\del}{\partial}
\newcommand{\be}{\begin{equation}}
\newcommand{\ee}{\end{equation}}
\newcommand{\ben}{\begin{eqnarray}}
\newcommand{\een}{\end{eqnarray}}
\newcommand{\nn}{\nonumber\\}
\def\ben{\begin{eqnarray}}
\def\een{\end{eqnarray}}
\def\be{\begin{equation}}
\def\ee{\end{equation}}

\begin{document}
\title{4D gravity on a non-BPS bent dilatonic brane}
\author{R.C. Fonseca$^{a}$, F.A. Brito$^{b}$, and L. Losano$^{a,c}$}
\affiliation{{\small {{$^a$}{Departamento de F\'\i sica, Universidade Federal da Para\'\i ba, 58051-970 Jo\~ao Pessoa PB, Brazil}}\\
{$^b$}{Departamento de F\'\i sica, Universidade Federal de Campina
Grande, \\Caixa Postal 10071, 58109-970 Campina Grande, Para\'\i ba, Brazil}}\\
{$^c$}{Centro de F\'\i sica do Porto, Rua do Campo Alegre 687, 4169-007 Porto, Portugal}}

\begin{abstract} 
 We investigate the localization of metastable four-dimensional gravity around a bent dilatonic brane, embedded into a five-dimensional space, that exists only up to distances sufficiently small compared to a crossover scale. Far from such scale, five-dimensional effects strongly deviate the Newtonian potential. We study this effect by considering localization of massive gravity on a non-BPS bent dilatonic 3-brane solution of a five-dimensional supergravity. Our results show that the cosmological constant on the bent brane controls the size of the crossover scale. For sufficiently small positive cosmological constant, that is in accord with the present observations, the crossover scale becomes very large.
\end{abstract}

\maketitle


We follow the original idea of Randall and Sundrum (RS) \cite{RS1}, to consider suitable extensions  to address the issue of localizing gravity in `bent' braneworld scenarios \cite{Kaloper:1999sm,Karch:2000ct}. In the RS scenario the five-dimensional gravity is coupled to a negative cosmological constant and a 3-brane sourced by a delta function. The solution in such setup is a symmetric solution given in terms of two copies of $AdS_5$ spaces patched together along the 3-brane. Although in this setup the fifth dimension is infinite the volume of the 5D bulk space is finite because the geometry is warped. As a consequence this allows having graviton zero mode responsible for 4D gravity on the brane. This is not necessary true for spaces whose volume diverges, because no zero mode emerges anymore. This was first shown by Gregory-Rubakov-Sibiryakov (GRS) \cite{RG} and Dvali-Gabadadze-Porrati (DGP) \cite{dvali}. The nice consequence of such an alternative setup is that 4D gravity on the 3-brane now emerges due to gravity massive modes and then is metastable. However, gravity massive modes can live long enough before escaping from the 3-brane to produce 4D gravity within a sufficient large scale  ---  the {\it crossover scale}.
In the present study we investigate such a scenario in a consistently truncated 5D supergravity \cite{susy_domain, Ref1}, where the 3-brane appears as a non-BPS bent dilatonic solution \cite{An}. They are made out of solutions of first order equations that also satisfy Einstein equations. We shall focus on the bosonic sector with 5D gravity coupled to two real scalar fields \cite{bbg, 027}.
In our investigations we are mainly interested on induced 4D gravity on specific bent dilatonic 3-brane solutions that can be embedded in an asymptotically 5D space with a diverging volume instead of asymmetric solutions \cite{An,Padilla:2004tp,CastilloFelisola:2004eg,Gabadadze:2006jm,FBL}. Bent dilatonic brane solutions have naturally appeared in the supergravity context in four- \cite{An} and five-dimensions \cite{Chamblin:1999ya, susy_domain, Ref1}.  Localization of gravity in such branes has also been addressed in \cite{Youm:2000xd,AlonsoAlberca:2000ff,Nojiri:2000yr} where finite volume and other issues were considered.  We shall focus on the infinite volume case, because they allow the possibility of metastable 4D gravity as first pointed out in the GRS \cite{RG} and DGP \cite{dvali} scenarios. Because the five-dimensional space has an asymptotically infinite volume then no gravity zero mode emerges. However, just as in GRS and DGP scenarios, we also have found 4D gravity that lives long enough within the crossover scale.  As we shall show below, the 4D gravity can last very much long as far as the cosmological constant on the brane is positive and very small, as currently observed. 
Thus, as emphasized in GRS and DGP scenarios, we shall focus on the main beautiful characteristic of the 4D metastable gravity, that is the fact that whereas gravity becomes four-dimensional for distances very much smaller than the crossover scale, it emerges as a five-dimensional gravity for distances very much larger than such scale. In doing so, we shall find the Newtonian potential induced by the gravity massive modes of a Schroedinger-like equation for the gravity fluctuations around the {\it bent dilatonic 3-brane solution}.

Le us consider the bosonic sector of the supergravity action for 
spacetimes in arbitrary $D$-dimensions $(D >3)$ coupled to ${\cal N}$ real scalar fields given by \cite{susy_domain,Ref1}
\be\label{s}
S=\int{d^Dx\;\sqrt{|g|}\left[\frac{1}{2\kappa^{D-2}}R-\frac12 g^{MN}\del_M\phi_i\del_N\phi_i -V(\phi_i)\right]},
\ee
where $\kappa = \frac1M_*$ is the $D$-dimensional Planck length, and the potential of the scalar fields are taken as
\be\label{v}
V(\phi_i)=2(D-2)^2\left[\left(\frac{\del W}{\del \phi_i}\right)^2-\kappa^{D-2}\left(\frac{D-1}{D-2}\right)W^2\right],
\ee
where $W(\phi_i)$ is the superpotential, and $\phi_i$, $i=1,2,..,{\cal N}$,  are real scalar fields.

Our present interest is looking for solutions from the equations of motion given by the action (\ref{s}) with potential (\ref{v}) describing braneworlds with cosmological constant. Furthermore we also concentrate our attention to localization of metastable gravity \cite{RG,dvali} that manifests extra-dimensional behavior at large distances and four-dimensional gravity at small distances. We end up concluding that the best solutions to gather these whole characteristics are bent dilatonic branes. In searching for such solutions let us first employ the `generalized Karch-Randall metric' \cite{Karch:2000ct}
\be\label{ref1}
ds^2_D = {\rm e}^{2\,A(y)}ds_{(D-1)}^2- dy^2,
\ee
where ${\rm e}^{A(y)}$ is the warp factor, $\mu,\, \nu = 0, 1, 2, ...,D - 2$ are indices on the $(D-2)$ -
brane. The line element of the bent four-dimensional space-time, $ds_{(D-1)}^2$, can have the general form 
\bes
\be\label{line}
ds_{(D-1)}^2=dt^2-{\rm e}^{\sqrt{\Lambda}\,t}\,(dx_1^2+dx_2^2+...+dx_{D-2}^2),
\ee
\be\label{line2}
ds_{(D-1)}^2={\rm e}^{-\sqrt{-\Lambda}\,x_{(D-2)}}\,(dt^2-dx_1^2- ... -dx_{(D-3)}^2)-dx^2_{(D-2)},
\ee
\ees
for dS geometry $(\Lambda>0)$ or AdS geometry $(\Lambda<0)$. By using the action \eqref{s} with the metric \eqref{ref1}, 
we obtain the set of
equations
\bes
\be\label{E.5a}
\phi_i''+(D-1)A'\phi_i'-\frac{\del V}{\del\phi_i}=0,
\ee
\be
A''= - \frac{\kappa^{D-2}}{(D-2)}\,\phi'^2-\Lambda\,{\rm e}^{-2\,A}
\ee
and
\be
A'^2 = \frac{2\,\kappa^{D-2}}{(D-1)(D-2)}\,\left(\frac12\,\phi'^2-V\right)+\Lambda\,{\rm e}^{-2\,A},
\ee
\ees
which are solved by the following first order equations
\bes
\be\label{ref1.2}
\del_yA = -\frac{2\kappa^{D-2}}{(D-1)(D-2)}\left(W+\Lambda\,\beta\,Z\right),
\ee 
\be\label{E.6b}\del_y\phi_i = \frac{3\,\kappa^{D-2}}{(D-1)(D-2)}\left(\frac{\del W}{\del\phi_i}+\Lambda(\alpha_i+\beta)\frac{\del Z}{\del\phi_i}\right),
\ee
\ees
where $Z=Z(\phi_i)$ \cite{027} is a new and in principle arbitrary function of the scalar field, to respond 
for the presence of the cosmological constant, $\alpha_i, \beta$ are real constants and we assume that the scalar fields only depend on the transverse coordinate $y$. 
We also make the transformation $W \rightarrow{W}/{[(D-1)(D-2)]}$, with $D=5$, in (\ref{v}) and use units which $\kappa^{D-2}=2$. This is to be in accord with the model presented in
Ref.\cite{027}.

The graviton modes on $(D - 2)$ - branes are governed by a linearized
gravity equation of motion in arbitrary number of
dimensions $(D>3)$ given by \cite{susy_domain, Ref1}
\be\label{r1.3}
\del_M(\sqrt{-g}g^{MN}\del_N\Phi) = 0,
\ee where $\Phi$ describes the wave function of the graviton on
non-compact coordinates $ M,N =0, 1, 2, ...,D - 1$. 
Let us consider $\Phi=h(y)\varphi(x^{\mu})$ into (\ref{r1.3}) and the fact
that $\Box_{D-1}\varphi =m^2\varphi$, where $\Box_{D-1}$ is the Laplacian operator
on the tangent frame. 

{ Before starting to compute the massive graviton wavefunction let us address the problem of the van Dam-Veltman-Zakharov (vDVZ) discontinuity. This problem states that in massive gravity at flat spacetime, the limit $m\to0$ cannot reduce the massive propagator to the massless one because extra longitudinal degrees of freedom persists. However, as it has been shown in \cite{Kogan:2000uy,Porrati:2000cp} this is not the case for massive gravity on `bent'  spaces such as $AdS_4$ and $dS_4$ spaces --- see also \cite{higuchi} for the $dS_4$ case only. This is precisely the situation of massive gravity on bent 3-branes (for $D=5$) presented above. To address the issue of vDVZ discontinuity one uses  a {\it ghost free} Pauli-Fierz massive gravity that in our case is just the induced massive gravity on the $D-2$-brane at $y=0$ (see paragraph with Eq.~(\ref{delta-source}) below)
\be\label{PFgravity}
S_{(D-1)}=\frac{1}{4}\int{d^{D-1}x\sqrt{{|\hat{g}|}}({\hat R}-\Lambda)}+\frac{1}{16}\int{d^{D-1}x\sqrt{|\bar{g}|}\,m^2(\bar{h}_{\mu\nu}^2-\bar{h}^2)},
\ee
where we have defined $ds^2_{(D-1)}=\bar{g}_{\mu\nu}dx^\mu dx^\nu$ as the metrics given in (\ref{line}) and (\ref{line2}) that corresponds to $(D-1)$ - dimensional Einstein space backgrounds  with 
$\bar{R}_{\mu\nu}=\frac{2\Lambda}{(D-3)} \bar{g}_{\mu\nu}$. $\hat{R}$ is made out of the metric defined as $\hat{g}_{\mu\nu}= \bar{g}_{\mu\nu} +  \bar{h}_{\mu\nu}$, where $\bar{h}_{\mu\nu}$ are the fluctuations around the background $\bar{g}_{\mu\nu}$. The four-dimensional linearized Einstein equations (for $D=5$) that follow from Eq.~(\ref{PFgravity}) by using {\it transverse traceless gauge} become $\square_{4}\bar{h}_{\mu\nu}+2\Lambda \bar{h}_{\mu\nu}=m^2 \bar{h}_{\mu\nu}$.  The suitable reduction of the massive propagator to the massless one is now controlled by the residue $\mbox{Res}=\frac{2\Lambda-2m^2}{3m^2-2\Lambda}$ at the physical pole \cite{Porrati:2000cp}.
This enables us to show \cite{Porrati:2000cp,Kogan:2000uy} 
that {\it there is no} vDVZ discontinuity for spaces with cosmological constant (positive or negative) as long as $m^2/\Lambda\to0$ as $m\to0$ such that Res $\to-1$ which corresponds to the structure of massless four-dimensional gravity. Notice that for $\Lambda=0$ (massive gravity in flat space) Res $\to-\frac23$ regardless the value of the mass. This confirms that in our setup with 3-brane developing massive gravity there is no vDVZ discontinuity since our 3-brane is bent by a nonzero cosmological constant. The vDVZ discontinuity is a pathology that plagues only Minkowski spaces as in DGP case since the tensorial structure of the propagators in their  four-dimensional massive gravity on the {\it flat} brane does not allow to recover the four-dimensional massless limit \cite{dvali}.
 }

Thus, the wave equation for
the graviton through the transverse coordinate $y$ reads
\be\label{ref1.4}
\frac{\del_y(\sqrt{-g}g^{yy}\del_y h(y))}
{\sqrt{-g}}
= -m^2|g^{00}|h(y).
\ee
This is our starting point to investigate graviton modes on the branes.
Using the components of the metric (\ref{ref1}) into the equation
(\ref{ref1.4}) we have
\be\label{ref1.5}
\frac12(D-1)\del_yA\del_yh(y) + \del_y^2h(y) = -m^2\,{\rm e}^{-2\,A(y)}h(y),
\ee
which can be recast in a more familiar equation by 
changing the metric (\ref{ref1}) in terms of the ``z-coordinate'' that leaves the metric in the conformally flat form
\be\label{ref1.6}
ds^2_D = {\rm e}^{2\,A(z)}(ds_{(D-1)}^2 - dz^2).
\ee
Now employing the changes of variables
$h(y)=\psi(z)\,{\rm e}^{-\frac{A(z)(D-2)}{2}}$ and $z(y)=
\int{
{\rm e}^{-{A(y)}}}dy$,
we can write the Schroedinger-like equation as follows
\be\label{ref1.7}
- \del^2 _z\psi(z) + V(z)\psi(z) = m^2\psi(z),
\ee
where the potential $V(z)$ is given by \cite{028}
\be\label{ref1.8}
V(z) =-\frac{(D - 2)^2}{4}\,\Lambda+
\frac{(D - 2)^2}{4}\,(\del_zA)^2 +\frac{D - 2}{2}\,\del^2_zA.\ee

Let us now examine the theory introduced by the action \eqref{s}, for five-dimensional gravity coupled to two real scalar fields, $\phi_1$ and $\phi_2$. The presence of the cosmological constant $\Lambda$ on the brane makes the problem much harder than the case of a Minkowski four-dimensional spacetime, which is obtained in the limit $\Lambda\rightarrow0$ --- this was already worked out in \cite{FBL}.
 In the case with cosmological constant on the brane, each term of the scalar potential (\ref{v}) is properly `shifted' by the $Z$-function so that for $D=5$ it is now given by \cite{027,028}
 \ben 
 V \left( \phi_{{1}},\phi_{{2}} \right) &=&\frac18\, \left( W_{{\phi_{{1}}}}+
\Lambda\, \left( \alpha_1+\beta \right) Z_{{\phi_{{1}}}} \right) 
 \left( W_{{\phi_{{1}}}}+\Lambda\, \left( -3\,\alpha_1+\beta \right) Z_
{{\phi_{{1}}}} \right) \nonumber 
\\ &  +&\frac18\, \left( W_{{\phi_{{2}}}}+\Lambda\,(\alpha_2+\beta)
\,Z_{{\phi_{{2}}}} \right)  \left( W_{{\phi_{{2}}}}+\Lambda\, \left( -
3\,\alpha_2+\beta \right) Z_{{\phi_{{2}}}} \right)-\frac13\, \left( W+
\Lambda\,\beta\,Z \right) ^{2},
 \een
and in order for the first order equations to satisfy the equations of motion we should impose the
following constraints
\bes \ben \label{ctr1}& &\alpha_1 W_{\phi_1} Z_{\phi_1\phi_1}+\alpha_1
Z_{\phi_1} W_{\phi_1\phi_1}+2\,\alpha_1\,\Lambda\,(\alpha_1+\beta)Z_{\phi_1}
Z_{\phi_1\phi_1}+\frac12(\alpha_1+\alpha_2)W_{\phi_2} Z_{\phi_1\phi_2}+\nn & &\alpha_2
Z_{\phi_2} W_{\phi_1\phi_2}+\frac12\Lambda\,(\alpha_2+\beta)\left(\alpha_1+3\,\alpha_2\right)
Z_{\phi_2} Z_{\phi_1\phi_2}-\frac43\,\alpha_1\, Z_{\phi_1}(W+\Lambda\beta Z)=0,
\\ & &  
\label{ctr2}
\alpha_2 W_{\phi_2} Z_{\phi_2\phi_2}+\alpha_2
Z_{\phi_2} W_{\phi_2\phi_2}+2\,\alpha_2\,\Lambda\,(\alpha_2+\beta)Z_{\phi_2}
Z_{\phi_2\phi_2}+\frac12(\alpha_1+\alpha_2)W_{\phi_1} Z_{\phi_1\phi_2}+\nn & &\alpha_1
Z_{\phi_1} W_{\phi_1\phi_2}+\frac12\,\Lambda\,(\alpha_1+\beta)\left(\alpha_2+3\,\alpha_1\right)
Z_{\phi_1} Z_{\phi_1\phi_2}-\frac43\,\alpha_2\, Z_{\phi_2}(W+\Lambda\beta Z)=0. \een \ees As solutions to the set of equations (\ref{E.5a})-(\ref{E.6b}) we use the {\it Ansatz} proposed in \cite{An}  to find {\it dilatonic  domain wall solutions}
 \bes
\be\label{An.a}
A \left( y \right) ={ {\ln  \left( 1-c_1
 y \right)  }}
,\quad y>0
\ee
\be\label{An.b}
A \left( y \right) ={ {\ln  \left( 1-{ c_2}\, y \right) }},\quad y<0
\ee
and
\be\label{An.c}
\phi_i(y)=-\frac{b_i}{b_1^2+b_2^2}\,A(y),
\ee
\ees
  We now chose a particular superpotential that satisfies the constraints (\ref{ctr1}) and (\ref{ctr2})
  \be
 W=-3\,a\,{{\rm e}^{b_1\phi_{{1}}+b_2\phi_{{2}}}}, 
 \ee
  where $a,b_1,b_2, c_1, c_2$ are real constants --- see also \cite{gonzalo} for other superpotentials. We also chose  a function $Z=W$, $\beta=0$, $c_i^2=a^2$, $\alpha_i=\alpha$, ${b_1^2+b_2^2}=\frac{2}{3}\frac{1}{1+\Lambda\alpha}$, such that the equations (\ref{E.5a})-(\ref{E.6b}) are satisfied. 
The orbits, $\phi_{{2}}={\frac {b_2\,\phi_{{1}}}{b_1}}$, in the scalar field space, decouple  the first-order equations and
give us BPS dilatonic  solutions. 
By using the proposed superpotential and this restrict set of parameters, the scalar potential is now given by
\be\label{PotSc}
V \left( \phi_1,\phi_2 \right) =-\frac94(1+\Lambda\alpha)a^2\,{{\rm e}^{2b_1\phi_{{1}}+2b_2\phi_{{2}}}}.\ee
The dilatonic solutions have the warp factors
 \bes
\be\label{An1.a}
A \left( y \right) = {\ln  \left( 1+{ c_1}\, y \right) }
,\quad y>0,
\ee
\be\label{An1.b}
A \left( y \right) = {\ln  \left( 1-{ c_1}\,  y \right) },\quad y<0,
\ee
and `kink' profiles
\be\label{An1.c}
\phi_i(y)=-\frac{b_i}{b_1^2+b_2^2}\, {\ln  \left( 1+{ c_1}\, y \right) }
,\quad y>0,
\ee
\be\label{An1.d}
\phi_i(y)=-\frac{b_i}{b_1^2+b_2^2}\, {\ln  \left( 1-{ c_1}\, y \right) }
,\quad y<0.
\ee
\ees
These solutions are also known as {\it scaling solutions} and have well discussed in \cite{thomas} and also very recently in \cite{thomas2}.
Although each `piece of solution' (for $y>0$ or $y<0$) is a BPS solution that also satisfies the equations of motion, we shall patch them together as follows
\be\label{An-T}
A \left( y \right) = {\ln  \left( 1+{ c_1}\, |y| \right) },
\qquad \phi_i(y)=-\frac{b_i}{b_1^2+b_2^2}\, {\ln  \left( 1+{ c_1}\,|y|\right) } ,
\ee
to address the issue of localizing gravity on the dilatonic brane. These latter solutions are non-BPS regular solutions, and they connect the same vacua $\Delta W\equiv W(\phi_i(+\infty)-W(\phi_i(-\infty))=0$. 
One should note that these solutions are consistent with delta function source on equations of motion since each piece of kink solutions can produce this term by taking the `thin wall limit' $b_i/(b_1^2+b_2^2)\to\infty$.

{Alternatively, this is equivalent to add a delta function at action (\ref{s}) to source the brane that will be responsible for the localization of gravity at $y=0$, i.e.,
\be\label{delta-source}
S\to S+\sigma\int{d^Dx\sqrt{|g|}\delta{(y)}},
\ee
where $\sigma=3c_1$ is a {\it positive} brane tension --- given in terms of $\frac32A'|_{y-\epsilon}^{y+\epsilon}$ with $\epsilon\to0$ \cite{susy_domain,An} . Notice this term does not contribute to the solutions  given by (\ref{An1.c}) and (\ref{An1.d}) for $y>0$ and $y<0$, respectively, of the original action (\ref{s}). However, it does affect the boundary conditions on the brane.
This has largely been adopted in four-dimensional dilatonic domain walls \cite{An} and embedded `delta function sourced' 3-branes in five dimensions \cite{susy_domain} --- see, e.g, the first reference. 

Furthermore, an effective negative bulk cosmological constant takes place around the brane, because the potential (\ref{PotSc}), written in terms of the superpotential, as a function of the kink profiles (\ref{An1.c}) and (\ref{An1.d}) approaches $ -\frac94(1+\Lambda\alpha)a^2$, where we assume $\alpha,\Lambda>0$, though no such bulk cosmological constant appears far ($y\to\pm\infty$) from the brane. Thus, both the brane tension and such an effective negative bulk cosmological constant around the brane play the role of localizing gravity on the brane.
As such, our setup develops a similar mechanism to localize gravity  as in both GRS and DGP scenarios with the exception that in GRS scenario \cite{RG} on also uses another brane with {\it negative tension} that is responsible for {\it ghosts}. }

In order to investigate the existence of gravitational modes in the vicinity of the branes we employ the aforementioned change of variable which results in the conformally flat metric (\ref{ref1.6})  to rewrite (\ref{An1.a}) and (\ref{An1.b}) to obtain the new form
\be\label{WP2}
A \left( z \right) =c_1\,|z|+1.
\ee  
For $D=5$, replacing (\ref{WP2}) in (\ref{ref1.8}) and (\ref{ref1.7}), we obtain the Schroedinger-like equation for the gravity fluctuations around the brane given by
\be\label{POT1}
-\frac{d^2}{d\,z^2}\psi_m\left( z \right) +\left(-\frac94\,\Lambda+\frac94\,{{ c_1}}^{2}+3\,c_1\,\delta \left( z \right)\right)\psi_m(z)
=m^2\psi_m\left( z \right),
  \ee where  $m^2$ is the four-dimensional mass which corresponds to the Kaluza-Klein modes. We have a Schroedinger problem for a modified delta potential barrier. For positive energies, the particles (e.g. bulk gravitons) are free to move in either half-space $z < 0$ or $z > 0$ and are scattered by the delta function potential at $z=0$. Since we are interested into the correction terms to the four-dimensional Newton law
between two unit masses on the brane, it is necessary to obtain the probability of
gravity with KK-modes on the brane. These modes are given in terms of the scattering states governed by $|\psi_m\left(0\right)|^2$ that depends on the magnitude of the transmission  $T$ or reflection $R$ coefficients. The jump condition at $z=0$,
\be\label{jump}
\left[{\frac {d}{dz}}\psi_{m} \left( z \right)\right]_{{z=0}}+ {3\,c_1} \, \psi_m \left( 0 \right) =0,
\ee
is obtained from the Schroedinger-like equation by using the properties of the delta function. Considering the general wave functions for scattered  states gives
\ben\label{wave}
\psi_{m1}(z)&=&\,{\rm{e}}^{{i\,\kappa\,z}}+R\,{\rm{e}}^{{-i\,\kappa\,z}}, \quad z<0;\nonumber\\
\psi_{m2}(z)&=&T\,{\rm{e}}^{i\,\kappa\,z}, \quad z>0.
\een where $\kappa^2=m^2+(9/4)\Lambda$ is the wave number.
 Then the probability density assumes the simplified form
\ben\label{Prob1}
\left| \psi_m \left( 0 \right)  \right|^2=|T|^2={\frac { \left( 9\,\Lambda+4\,{m}^{2} \right) {k}^{2}}{ 
1+ \left( 9\,\Lambda+4\,{m}^{2} \right) {k}^{2} }},
\een where $k^2=\frac{1}{9\,c_1^2}$ and we have absorbed the second term of the Schroedinger potential into the cosmological constant $\Lambda$-term.
The correction to the four-dimensional Newtonian potential
generated by the massive modes is given by \cite{RS1}
\be\label{03}
V (r) = \frac{M^{-3}_{5}}{r}|\psi_0(0)|^2 + \Delta V (r) ,
\ee 
where the first term is contribution of zero mode and the second term corresponds to
the correction term which is generated by the exchange of KK-modes. Since our problem has no zero mode, then 
\be\label{04}
V(r)=\Delta V (r) = \frac{M^{-3}_{5}}{r}\int^{\infty}_{0}{dm\,{\rm e}^{-m\,r}|\psi_m(0)|^2}.
\ee Our analysis takes place in a single region, where we obtain the probabilities for existence of gravity with continuous mode on the brane at
$z=0$.
Substituting $(\ref{Prob1})$ in (\ref{04}), we can get approximately the form of $\Delta V (r)$ at a distance $r$. By using the relation ${\rm Ei}(i\,x)={\rm ci}(x)+i\left[1/2\,\pi+{\rm si}(x)\right]$, where ${\rm ci}(x)=-\int_x^{\infty}{\cos\,t/t}=\gamma+\ln(x)+\rm{Ci}(x)$ and ${\rm si}(x)=-\int_x^{\infty}{\sin\,t/t}=\rm{Si}(x)-\frac{\pi}{2}$, the dominant term (as $r$ becomes large but is still much smaller than the crossover scale) is
\ben\label{CiSi}
V \left( r \right) = \frac{M^{-3}_5}{4\,r\,f(k)\,k}\, \left( \sin \left( \frac12\,{\frac {f \left( 
k \right) r}{k}} \right) {\rm Ci} \left( \frac12\,{\frac {f \left( 
k \right) r}{k}} \right) -\cos \left( \frac12\,{\frac {f \left( 
k \right) r}{k}} \right) {\rm Si} \left( \frac12\,{\frac {f \left( 
k \right) r}{k}} \right)  \right)   ,
 \een 
where $f \left( k \right) =\sqrt{\Lambda\,k^2+1}$. Let us now discuss the large and small distance behavior comparing with the crossover scale $r_c=k/f(k)$. { For flat branes $f(k)\to1$ and then $r_c\to k$ since $r_c\sim 1/m$ we should also have $k\sim 1/m$. As we turn on the cosmological constant, as in the case of bent branes, we find $r_c=k/f(k)\sim 1/
\sqrt{\Lambda+m^2}$.  This properly gives $r_c\sim 1/m$ as $\Lambda\to0$ and $r_c\sim 1/\sqrt{\Lambda}$ as $m^2<<\Lambda$ with $\Lambda>0$, $m\sim 1/k$. The latter  limit is phenomenological interesting and clearly avoid the vDVZ discontinuity earlier discussed. 
} Notice this asks a positive cosmological constant on the brane. In the following we shall use the asymptotic forms: ${\rm Ci}(x)\sim\gamma+\ln(x)$ and ${\rm Si}(x)\sim x-{\pi}/{2}$ for $x<<1$, $
{\rm Ci}(x)\sim{{\rm sin}}(x)/{x}$ and ${\rm Si}(x)\sim -{{\rm cos}}(x)/{x}$ for $ x>>1$. 

For small distance, i.e., $r/r_c<<1$,
where $\gamma\approx0.577$ is the Euler-Masceroni constant, we obtain the following form 
\be\label{13}
V \left( r \right) \sim \frac{M^{-3}_5}{4\,r\,f(k)\,k}\,\left(\frac12\,\frac{f \left( k \right) r}{k} \left( \gamma+\ln  \left( \frac12\,{\frac 
{f \left( k \right) r}{k}} \right)  \right) -\frac12\,{
\frac {f \left( k \right) r}{k}}+\frac12\,\pi+{\cal O}\left(r^2\right)\right),
\quad r<<{r_c}.
\ee
As we expected, at short distances the potential has the correct $4D$ Newtonian $1/r$
scaling. This is subsequently modified by the logarithmic repulsion term in (\ref{13}).

Finally, for large distance, i.e., $r/r_c>>1$, the potential in Eq.~(\ref{CiSi}) gives
 \be\label{15}
V(r)\sim \frac{\,M_5^{-3}}{2\,f \left( k \right)^2{r}^{2}}\sim \frac{1}{r^2}
,\quad\quad r>>{r_c}, 
\ee
in accordance with the laws of $5D$ gravity \cite{RG,dvali}. {For a recent comprehensive review on modified gravity and cosmology see \cite{tc2011}. }
\\


In summary, it is shown that the gravitational potential becomes
the usual Newton law $(\rightarrow 1/r)$ at short distance and five-dimensional law $(\rightarrow 1/r^2)$ at large distance.
This study showed that from a $5D$ supergravity theory with two scalar fields with standard dynamics the emergence of $4D$ gravity on a non-BPS bent dilatonic  brane exists even for an asymptotic $5D$ space with infinite volume, below a crossover scale,  and the manifestation of extra dimensions does not necessarily occur only at short distances as commonly expected. The complete
behavior is controlled by a crossover scale $r_c$ --- see \cite{023} for similar results and \cite{Neupane:2010ey,Wang:2002pka,Cruz:2010zz, Landim:2011ki} for related issues. This is also in accord with GRS and DGP scenarios, and now we have shown that $r_c$ is indeed related to the Hubble size given in terms of the cosmological constant on the brane, i.e. $r_c\sim 1/\sqrt{\Lambda}$ which implies a positive cosmological constant what is in accord with the present observations. The $4D$ gravity can live long enough on the brane as far as such a positive cosmological constant is very small. {
}

The authors would like to thanks CAPES,
CNPq, CAPES/PROCAD/PNPD and PRONEX/CNPq/FAPESQ for partial support.

\end{document}